# Upper critical field, critical current density and activation energy of the new La$_{1-x}$Sm$_x$O$_{0.5}$F$_{0.5}$BiS$_2$ (*x* = 0.2, 0.8) superconductors

G. Kalai Selvan[1], G. S. Thakur[2], K. Manikandan[1], Y. Uwatoko[4], Zeba Haque[2], L. C. Gupta[2,†], A. K. Ganguli[2,3,*] and S. Arumugam[1,*]

[1]Centre for High Pressure Research, School of Physics, Bharathidasan University, Tiruchirapalli 620024, India.

[2]Department of Chemistry, Indian Institute of Technology New Delhi 110016, India.

[3]Institute of Nano Science and Technology, Mohali, 160062, India.

[4]Institute of Solid State Physics, University of Tokyo, 5-1-5 Kashiwanoha, Kashiwa, Chiba 277-8581, Japan.

**Abstract**

Critical current density ($J_c$), thermal activation energy ($U_0$), and upper critical field ($H_{c2}$) of La$_{1-x}$Sm$_x$O$_{0.5}$F$_{0.5}$BiS$_2$ (*x* = 0.2, 0.8) superconductors are investigated from magnetic field dependent $\rho(T)$ studies. The estimated upper critical field ($H_{c2}$) has low values of 1.04 T for *x* = 0.2 and 1.41 T for *x* = 0.8. These values are lower than Sm free LaO$_{0.5}$F$_{0.5}$BiS$_2$ superconductor (1.9 T). The critical current density ($J_c$) is estimated to be 1.35*10$^5$ A/cm$^2$ and 5.07*10$^5$ A/cm$^2$ (2 K) for *x* = 0.2 and 0.8 respectively, using the Bean's model. The thermal activation energy ($U_0/k_B$) is 61 K for *x* = 0.2 and 140 K for *x* =0.8 as calculated from Arrhenius plots at low magnetic field (1 T) and indicates a strong flu*x* pinning potential which might be co-existing with applied magnetic field.

===============================================================================

**\*Corresponding Authors:**

**Prof. S. Arumugam:** Email:-**sarumugam1963@yahoo.com**; Telephone no: - (O): 91-431-2407118; Fa*x* No: - 91-431- 2407045, 2407032

**Prof. A. K. Ganguli :** Email :- **ashokganguliiitd@gmail.com**; Phone Numbers : +91-172-2210075/57/56, Fax Number: +91-172-2211074

===============================================================================

**Introduction**

Recent discovery of BiS$_2$-based superconductors has attracted considerable attention [1–12]. The first BiS$_2$-based superconductor discovered with superconducting transition temperature $T_c \sim$ 4.5 K, was Bi$_4$O$_4$S$_3$, which is composed of Bi$_2$O$_2$(SO$_4$)$_{0.5}$ blocking layers and BiS$_2$ superconducting layers [13]. Subsequently, a series of BiS$_2$-based superconductors has been synthesized with $T_c$ of 2 - 10 K for ReO$_{1-x}$F$_x$BiS$_2$ (Re = La, Ce, Pr, Nd and Yb) [1,2,6,7,14], La$_{1-x}$M$_x$OBiS$_2$ (M = Ti, Zr, Hf, Th) [4], Sr$_{1-x}$Ln$_x$FBiS$_2$ [9,8,15]. $T_c$ in these materials could be enhanced by chemical pressure or external pressure [16–25]. Among all these materials, LaO$_{0.5}$F$_{0.5}$BiS$_2$ exhibits the highest onset $T_c \sim$ 10.6 K [1] when synthesized under high pressure. Since the highest $T_c$ in the LaO$_{0.5}$F$_{0.5}$BiS$_2$ samples grown under ambient pressure is ~2.8 K [1], we systematically substituted Sm at La sites in LaO$_{0.5}$F$_{0.5}$BiS$_2$ and observed an enhancement in $T_c$ upto ~ 5 K with 80% Sm substitution [24]. It may be noted that full Sm-based compound SmO$_{1-x}$F$_x$BiS$_2$ compound doesn't exhibit superconductivity [24], though the smaller rare-earth ion Yb-based compound is superconducting [6]. The crystal structure of Ln(O,F)BiS$_2$ consists of a charge reservoir layer (*Ln*O) alternated with BiS$_2$ planes that play the same role as CuO$_2$ planes in cuprates or the FeAs and Fe(Se,Te) planes in iron-based superconductors (IBSs). However, contrary to both cuprates and IBSs, where the conduction bands at the Fermi energy are dominated by 3$d$ states, in BiS$_2$ systems the main contribution comes from the bismuth 6$p$ orbitals, which are spatially extended and hybridized with the 3$p$ orbitals of sulfur [26–28].

Upper critical field $H_{c2}$ and the critical current density $J_c$ are the two characteristic properties that are important while considering the application potential of a superconductor. Mizuguchi *et al.*, reported that the first BiS$_2$ based compounds show an upper critical field of 1.5 T in Bi$_4$O$_4$S$_3$ ($T_c \sim$ 4.5 K) and 1.2 T in CeO$_{0.5}$F$_{0.5}$BiS$_2$ ($T_c \sim$ 2.5 K) [6,13]. $H_{c2}$ and $T_c$ could be enhanced up to ~10 T and 10 K on replacing Ce by La, Nd, Pr and/or synthesizing LaO$_{0.5}$F$_{0.5}$BiS$_2$ under high pressure [1,3,6,7]. Several investigations have already shown that the value of $J_c$ is higher under very low magnetic field and with high transition temperature ($T_c$) as for example in Fe-based superconductors such as SmFeAsO$_{0.7}$F$_{0.3}$ and NdFeAsO$_{0.7}$ ($J_c \sim$ 5*10$^5$ A/cm$^2$) [29]. Till date critical current density ($J_c$) has not been reported in any of the BiS$_2$ based superconductors.

Recently, Liu et al. reported that the single crystal of $NdO_{0.5}F_{0.5}BiS_2$ has the highest $H_{c2}$ up to 40 T which is the maximum $H_{c2}$ observed so far at very low temperature in $BiS_2$ based superconductors [30]. It is well known that the normal state magnetic and transport properties can provide the crucial information about the fundamental parameters of a superconducting material. In this paper, we extend our previous work on the $La_{1-x}Sm_xO_{0.5}F_{0.5}BiS_2$ ($x$ = 0.2 and 0.8) superconductors to investigate the important fundamental parameters ($H_{c2}$, $J_c$, $\xi_{GL}$ and $U_0$) by using Werthamer-Helfand-Hohenberg (*WHH*) formula, scanning electron microscopic (*SEM*) studies, Beans model and Arrhenius plots.

**Experimental details**

Polycrystalline samples with the nominal compositions of $La_{1-x}Sm_xO_{0.5}F_{0.5}BiS_2$ ($x$ =0.2 and 0.8) were synthesized by solid state reaction and characterized by powder X-ray diffraction technique as reported earlier. Field dependent magnetization and resistivity measurements were carried out using Physical Property Measurement System (*PPMS*)-Vibrating Sample Magnetometer (*VSM*) (Quantum Design, USA). Microstructure analyses of polycrystalline samples were carried out using Scanning Electron Microscopy (*SEM*) with operated at a voltage of 20 kV. The field dependent upper critical field ($H_{c2}$) and critical current density ($J_c$) was calculated using Werthamer-Helfand-Hohenberg (*WHH*) formula and Bean's model respectively. Thermal activation energy ($U_0$) was calculated from slope of the linear part of temperature dependent Arrhenius plots.

**Results and Discussion**

All the synthesized samples crystallize in a tetragonal *P4/nmm* space group. Few small peaks for impurity phases like $Bi_2S_3$ (~3 %) and $LaF_3$ (~3 %) for $x$ = 0.2 and $Bi_2S_3$ (~8 %) for $x$ = 0.8 composition were detected in X-ray analysis [24]. $La_{1-x}Sm_xO_{0.5}F_{0.5}BiS_2$ showed $T_c$ ~ 3.2 K (for $x$ = 0.2) and ~ 4.6 K (for $x$ = 0.8) which are higher than the parent compound $LaO_{0.5}F_{0.5}BiS_2$ (2.7 K). This possibly could be due to increase in chemical pressure brought by substitution of smaller Sm atoms at the La site. Figure.1 shows the magneto-resistance plots for $La_{1-x}Sm_xO_{0.5}F_{0.5}BiS_2$ ($x$ = 0.2 and 0.8) as a function of temperature, which gives the details of upper critical field $H_{c2}$ and flux pinning properties in magnetic field up to 1 T. $H_{c2}$ and the irreversibility field $H^*$ were calculated using the criteria of 90% and 10% of normal state resistivity ($\rho_n$). The superconducting transition width is determined as 0.2 K [Fig.1 (a)] and 0.15 K [Fig.1 (b)] and sharpness of the transition indicate high crystallinity and good structural homogeneity of the sample.

It can be seen that both the $T_c^{onset}$ and the $T_c^{zero}$ shift rapidly towards lower temperature with magnetic field as shown in Fig.1 (a, b). The conventional one-band Werthamer–Helfand–Hohenberg (*WHH*) equation ($H_{c2}(0) = -0.693T_c(dH_{c2}/dT)T=T_c$) is used to determine the upper critical field [31]. Using $H_{c2}$ versus $T_c$ plot (inset of Fig. 1 (a,b)), the slope of ($dH_{c2}/dT_c$) was calculated as -0.476 [T/K] and -0.446 [T/K] for $La_{0.8}Sm_{0.2}O_{0.5}F_{0.5}BiS_2$ and $La_{0.2}Sm_{0.8}O_{0.5}F_{0.5}BiS_2$ respectively. $H_{c2}(0)$ was found to be 1.04 T and 1.41 T for $x = 0.2$ composition (with $T_c = 3.2$ K) and $x = 0.8$ (with $T_c = 4.6$ K), which is slightly lower than reported for Sm free $LaO_{0.5}F_{0.5}BiS_2$ superconductor ($H_{c2} \sim 1.9$ T) [32]. It is clear from Fig.1 (a,b) that the broadening of $d\rho/dT$ peak (not shown in fig) takes place with increase in the applied magnetic field, which suggests that the $T_c$ onset is relatively less affected than the $T_c(\rho=0)$. The sharp drop in resistivity near the superconducting transition region gives narrow transition width, centered at mid of $T_c$ in zero field, indicating good percolation path of superconducting grains, which is an indication towards the better granular coupling in this compound. The broadening of the resistive transitions in the $\rho(T)H$ curve with increasing magnetic fields can be seen in terms of energy dissipation due to vortex motion [33]. From the resistivity measurements, the calculated residual resistivity ratio ($RRR=R_{300K}/R_{11K}$) above 11 K was found to be 0.50 ($x = 0.2$) and 0.73 ($x = 0.8$). From this, the Ginzburg-Landau coherence length ($\xi$) = $[\Phi_0/(2\pi H_{c2})]^{1/2}$, where $\Phi_0 = 2.07*10^{-7}$ Gcm$^2$ the zero-temperature coherence length $\xi_{GL}(0)$ was estimated to be ~176 Å for $La_{0.2}Sm_{0.8}O_{0.5}F_{0.5}BiS_2$ and ~158 Å for $La_{0.8}Sm_{0.2}O_{0.5}F_{0.5}BiS_2$ which are slightly higher than the values reported for Sm-free $LaO_{0.5}F_{0.5}BiS_2$ ($\xi \sim 130$ Å)[32]. Further, as seen in both the Figures 1(a,b), we note that bulk superconductivity is suppressed with application of a small magnetic field greater 0.5 T, as shown in $\rho(T)$ curves. In an earlier study, this behavior has been interpreted as residual Cooper pairs existing in the system even when the bulk superconductivity has been completely suppressed, which supports the applicability of strong coupling theories [34].

To investigate the microstructure and dispersion of superconducting grains in polycrystalline $La_{1-x}Sm_xO_{0.5}F_{0.5}BiS_2$ ($x$ = 0.2, 0.8) SEM analysis was performed which are shown in figure 2 (a, b). The formation of agglomerated plate-like structures was evident in $La_{1-x}Sm_xO_{0.5}F_{0.5}BiS_2$ ($x$ = 0.2, 0.8). The average grain size calculated from *SEM* images through *Image j* software was found to be ~ 0.73 $\mu m$ ($x$ = 0.2) and ~ 0.56 $\mu m$ ($x$ = 0.8) which is illustrated in Fig. 2 (c, d), and found that, the $x$ = 0.8 has rather smaller grains compared to $x$ = 0.2. These results reveal that the chemical pressure generated by Sm substitution at La sites reduces the crystallites size. The figure 3 (a,b) shows the magnetic field dependence of critical current density at 2 K. Using Bean's model [35] the critical current density was calculated from *M(B)* curves (shown in the insets of figure 3 a and b), using the formula, $J_c = 30\Delta M/d$, where $\Delta M$ is the difference in decreasing and increasing branch of magnetization (($M_{dec} - M_{inc}$)/2) under applied magnetic field (in Oe) and d is the average grain size 0.73 $\mu m$ ($x$ = 0.8) and 0.56 $\mu m$ ($x$ = 0.8) superconductors. Insert shows the field dependence of magnetization of $La_{1-x}Sm_xO_{0.5}F_{0.5}BiS_2$ ($x$ = 0.2, 0.8) superconductors at low temperatures (2 K). The lower critical field ($H_{c1}$) can be defined as the deviation from the linear *M(H)* curve as shown in the same figure. $H_{c1}$ is found to be ~39 Oe for $x$ = 0.8 and ~ 47 Oe for $x$ = 0.2 at 2 K, which is higher than reported in other $BiS_2$ based systems such as $PrO_{0.5}F_{0.5}BiS_2$ (8 Oe) [7] and $NdO_{0.5}F_{0.5}BiS_2$ (25 Oe) [3]. The calculated $J_c$ value for the $x$ = 0.2 sample was found to be 1.35*10$^5$ A/cm$^2$ at low magnetic field (50 Oe). Further, there is no peak effect observed at 2 K up to 400 Oe in $x$ = 0.2 sample. $J_c$ for $x$ = 0.8 sample is found to be 5.07*10$^5$ A/cm$^2$, which is slightly higher than the value for $x$ = 0.2 sample indicating a strong flu$x$ pinning force, arising due to Sm doping. Further, the $J_c$ decreases monotonically with increasing applied magnetic field up to 300 Oe for both $x$ = 0.2 and 0.8, which is similar to field dependence of $J_c$ in LnOFeAs (Ln = Nd and Sm) compounds [36,37]. Magnitude of $J_c$ comparable with reported previously FeAs and cuprate superconductors [33,38]. It should be noted that the low magnitude of $J_c$ is very likely caused by the interface between grains. As the field increases, these interfaces act as weak-links and dramatically reduce $J_c$. As $J_c$ is determined from the magnetization experiment, the small hysteresis comes from individual grains which are mostly disconnected. Therefore, it is e$x$pected that the real $J_c$ of individual grains should be higher than the $J_c$ calculated using real sample dimensions.

To test the potential of BiS$_2$ based superconductors for applications, we discuss the basic pinning and vortex activation characteristics of La$_{1-x}$Sm$_x$O$_{0.5}$F$_{0.5}$BiS$_2$ ($x$ = 0.2, 0.8). The broadening of the resistive transition under an applied magnetic field is understood in terms of the dissipation of energy caused by the motion of vortices in the mixed state. This interpretation is based on the fact that for the low resistance region, the loss is caused by the creep of vortices, which is modelled as Arrhenius type thermally activated behavior. This is simply described by $\rho(T, B)=\rho_0 \exp(-U_0/k_BT)$, where $U_0$ is flux flow energy or pinning potential, $\rho$ and $\rho_0$ are the normal state and residual resistivity, and $k_B$ is the Boltzmann constant (1.380*10$^{-23}$m$^2$ kg s$^{-2}$). In Fig. 4 (a,b), we have shown the Arrhenius plot ($ln(\rho)$ vs $T^{-1}$) and it could be clearly seen that $U_0$ is strongly combined with applied magnetic field. The activation energy $U_0$ can be calculated from the slope of the line in Arrhenius plot. The magnetic field dependence of the activation energy ($U_0$) is shown in the insets of Fig. 4 (a,b). The best fit to the experimental data yield the values of the activation energy $U_0/k_B$ = 140 K and 61 K in zero-field for the $x$ = 0.8, 0.2 samarium doped samples, respectively. Srivastava *et al*, recently reported that a decrease in the activation energy ($U_0/k_B$) from ~121 to 9.75 K at 0.01 T to 1 T was found in Bi$_4$O$_4$S$_3$[39]. We can see that the $U_0/k_B$ for both samples is comparable for a field of 1 T and a similar trend was observed for Sm = 0.2 sample. Both the figures in inset shows gradual decrease in $U_0$ with increase in magnetic field (from $H$ = 0 to 1 T) for the Sm doped superconductor, which might be caused by strong vortex pinning forces co-existing with the applied magnetic field.

**Conclusions**

We have studied the upper critical field, critical current density, microstructure, and thermally activated flux energy of the La$_{1-x}$Sm$_x$O$_{0.5}$F$_{0.5}$BiS$_2$ ($x$ = 0.2, 0.8) superconductors. Temperature dependent magneto- resistance studies show a slight improvement of upper critical field in Sm rich sample ($H_{c2}(0)$ ~ 1.04 T (for $x$ = 0.2) and 1.41 T ($x$ = 0.8)). For the first time the critical current density has been reported in Ln(O,F)BiS$_2$ superconductors. A high $J_c$ value (5.07 *10$^5$ A/cm$^2$) was found for La$_{0.2}$Sm$_{0.8}$O$_{0.5}$F$_{0.5}$BiS$_2$ at 2 K which is comparable to the reported Fe based oxypnictides. The estimated thermal activation energy $U_0/k_B$ ~ 140 K (at 0 T) for La$_{0.2}$Sm$_{0.8}$O$_{0.5}$F$_{0.5}$BiS$_2$ sample is slightly higher than La$_{0.8}$Sm$_{0.2}$O$_{0.5}$F$_{0.5}$BiS$_2$ (La – rich) superconductor and those previously reported for Ln(O,F)BiS$_2$ (Ln = Ce, Pr, Nd, and Yb) superconductors. Our studies show that on substitution of smaller rare earth metal (Sm) in place of La in $LaO_{0.5}F_{0.5}BiS_2$ successfully improves and enhances magnetic flux pinning forces making this superconductor a potential candidate for superconducting applications.

## Acknowledgements

The author SA wishes to thank DST, (SERB, TSDP, FIST) BRNS, DRDO, CEFIPRA and UGC for the financial support. GKS would like to thank UGC- BSR- RFSMS- SRF for the meritorious fellowship. AKG thanks Department of Science & Technology, Govt. of India for funding the project on superconducting materials. GST thanks CSIR, Govt. of India for a fellowship.

## Notes and References

**Figure captions**

**Figure.1.** Magnetic field dependence of resistivity as a function of temperature for (a) $La_{0.8}Sm_{0.2}O_{0.5}F_{0.5}BiS_2$ and (b) $La_{0.2}Sm_{0.8}O_{0.5}F_{0.5}BiS_2$ superconductors. Inset shows temperature dependence of upper critical field (♦) and irreversibility field (●).

**Figure.2.** SEM micrographs of (a) $La_{0.8}Sm_{0.2}O_{0.5}F_{0.5}BiS_2$ and (b) $La_{0.2}Sm_{0.8}O_{0.5}F_{0.5}BiS_2$ superconducting grains within the range of 0.73 $\mu m$, 0.56 $\mu m$. (a1) and (b1) show the particles distribution of the same samples.

**Figure.3.** Illustrates magnetic field dependence of critical current density ($J_c$) at 2 K. Inset shows magnetic hysteresis loops for (a) $La_{0.8}Sm_{0.2}O_{0.5}F_{0.5}BiS_2$ and (b) $La_{0.2}Sm_{0.8}O_{0.5}F_{0.5}BiS_2$ superconductors measured at 2 K, arrow marks indicate the lower critical field ($H_{c1}$).

**Figure.4.** Arrhenius plots of normal state resistivity of (a) $La_{0.8}Sm_{0.2}O_{0.5}F_{0.5}BiS_2$ and (b) $La_{0.2}Sm_{0.8}O_{0.5}F_{0.5}BiS_2$ with H= 0 to 1 T. The activation energy ($U_o/k_B$) at an applied magnetic field is obtained by the slope of the linear fit. Inset shows magnetic field dependence of the activation energy ($U_o/k_B$).

**Figure. 1**

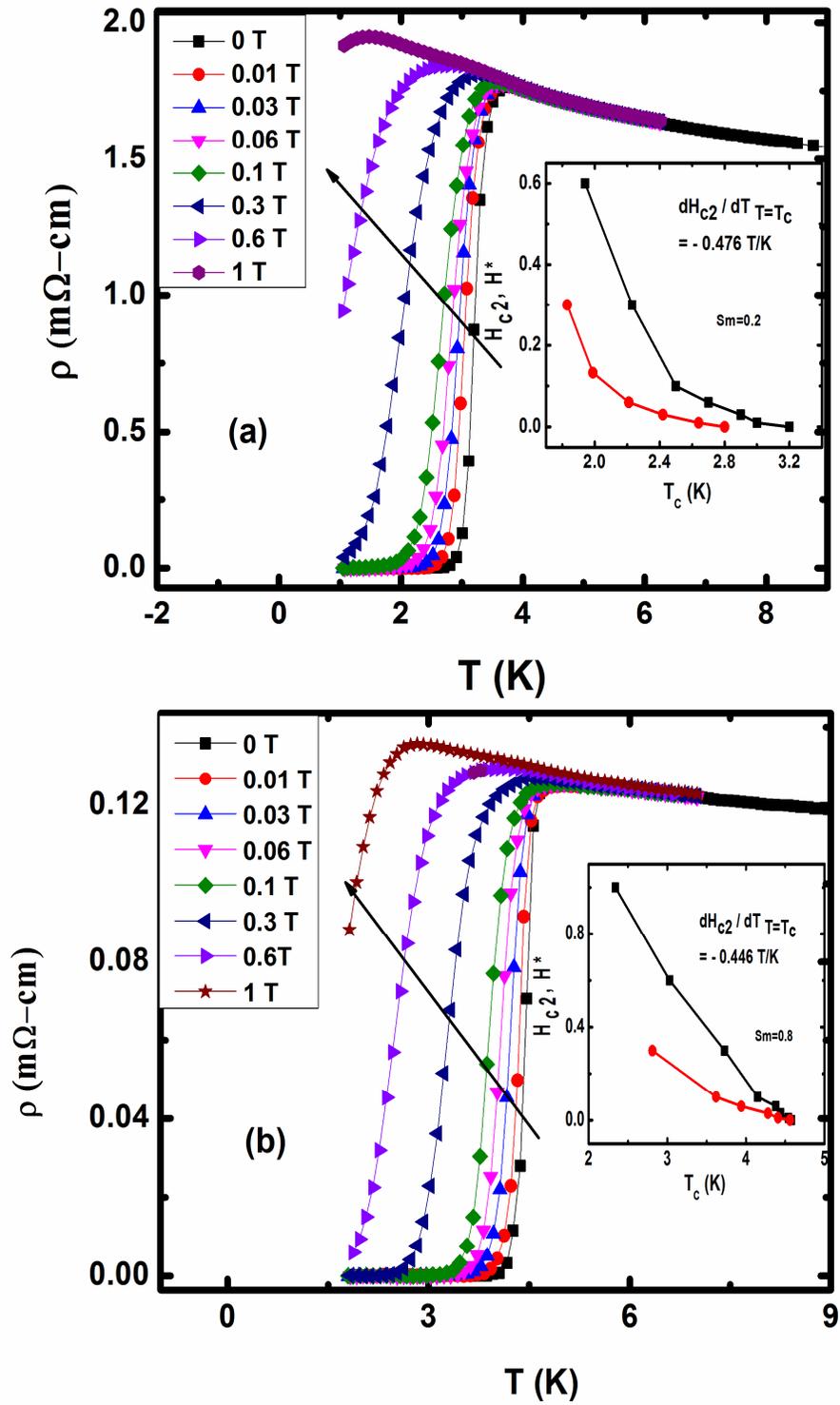

**Figure. 2**

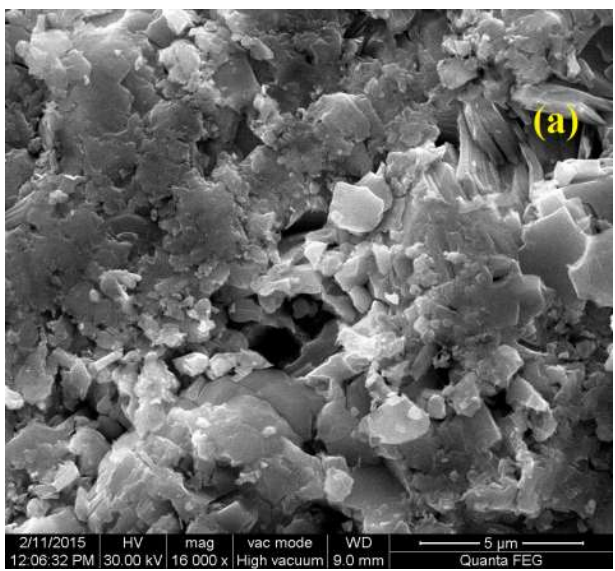
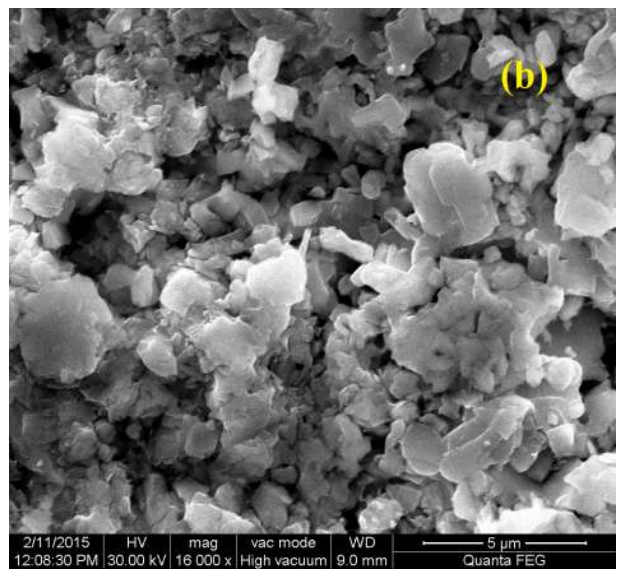
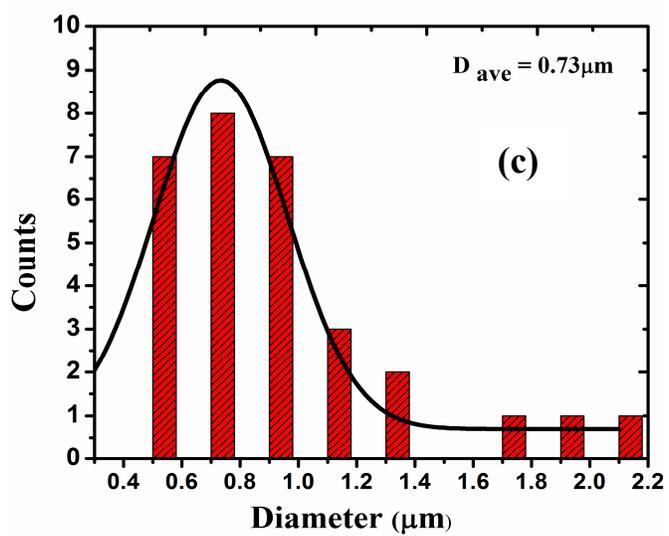
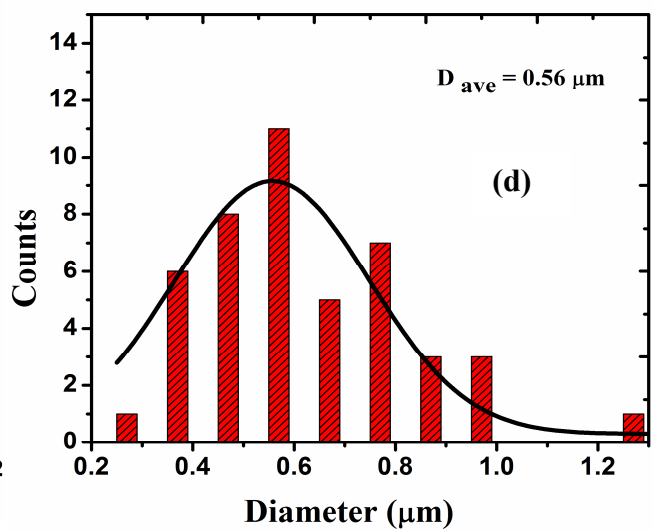

Figure. 3

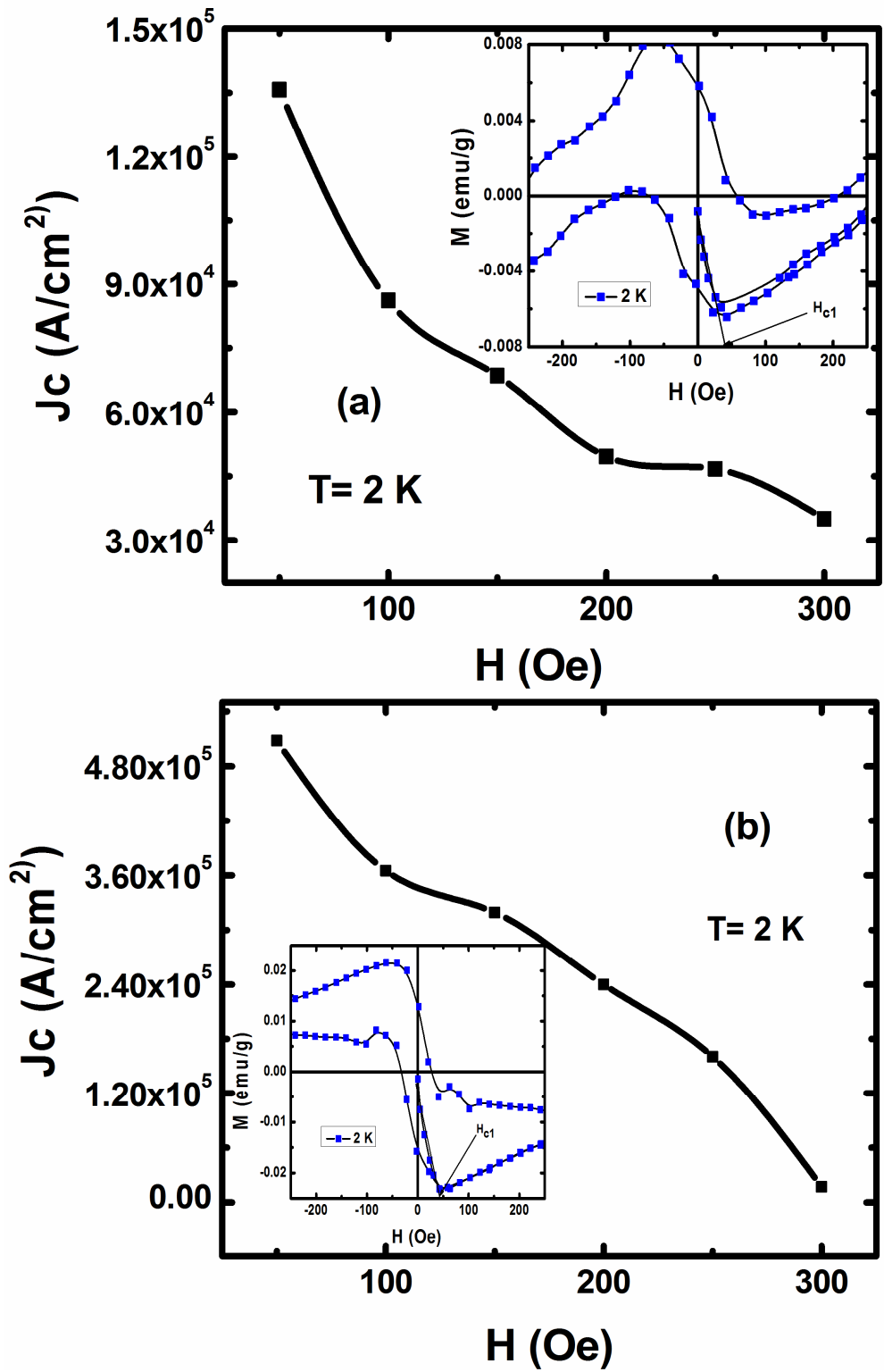

**Figure. 4**

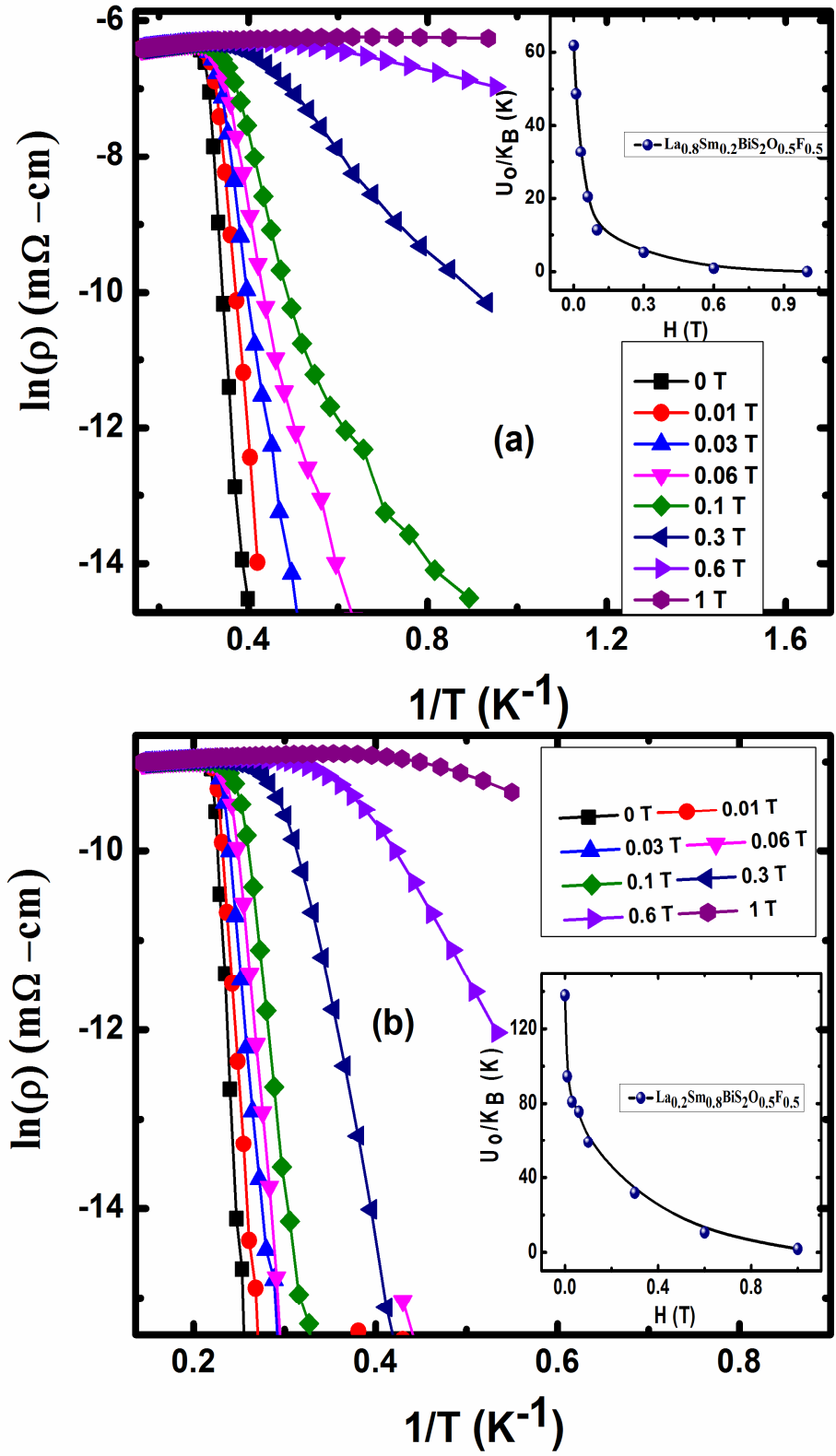